\documentclass[twocolumn]{jpsj3}
\usepackage{txfonts}
\usepackage{times}
\usepackage{color}

\title{Superconductivity and Ferromagnetic Quantum Criticality in Uranium Compounds}

\author{
Dai Aoki$^{1,2}$\thanks{E-mail address: aoki@imr.tohoku.ac.jp}, 
and
Jacques Flouquet$^2$
}

\inst{
$^1$IMR, Tohoku University, Oarai, Ibaraki, 311-1313, Japan\\
$^2$INAC/SPSMS, CEA Grenoble, 38054 Grenoble, France\\
}

\abst{
We review our recent studies on ferromagnetic superconductors, UGe$_2$, URhGe and UCoGe, together with the ferromagnetic quantum criticality and paramagnetic singularity on the Ising 5$f$-itinerant system UCoAl.
Thanks to the variety of ordered moment in ferromagnetic superconductors from $1.5\,\mu_{\rm B}$ to $0.05\,\mu_{\rm B}$,
interesting systematic changes or similarities are clarified.
All ferromagnetic superconductors show large upper critical field $H_{\rm c2}$,
and the field-reentrant (-reinforced) phenomena are observed in the field-temperature phase diagram,
when the pressure or field direction is tuned for particular conditions.
These phenomena are well explained by the ferromagnetic longitudinal fluctuations,
which are induced by the magnetic field in transverse configurations.
The large $H_{\rm c2}$ might be also associated with possible additional effects of Fermi surface instabilities, such as Lifshitz-type singularities.
}

\begin{document}
\maketitle

\section{Introduction}
Ferromagnetism and superconductivity had been thought to be antagonistic, because the large internal field due to the ferromagnetism easily destroys the Cooper pairs for conventional $s$-wave superconductors.
Nevertheless some materials, such as ErRh$_4$B$_4$~\cite{Fer77}, HoMo$_6$S$_8$~\cite{Ish77},
show superconductivity and ferromagnetism, but no coexistence.
The Curie temperature $T_{\rm Curie}$ is lower than the superconducting critical temperature $T_{\rm sc}$, and the two orders are considered to be competing.

The coexistence of ferromagnetism and superconductivity
was discovered for the first time in UGe$_2$ under pressure near the critical pressure $P_{\rm c}$ of ferromagnetism~\cite{Sax00}.
Soon after, the superconductivity was found in the ferromagnet URhGe at ambient pressure~\cite{Aok01}.
More recently, UCoGe was reported as a new member of ferromagnetic superconductors~\cite{Huy07}.
All three materials show the microscopic coexistence of ferromagnetism and superconductivity proved by NMR/NQR, neutron scattering and $\mu$SR experiments~\cite{Kot05,Hux03,Oht08,Hat13,Vis09}, 
and $T_{\rm sc}$ is lower than $T_{\rm Curie}$.
The ordered moments of uranium are much lower than the expected free ion values.
The 5$f$ electrons are considered to be itinerant, and they contribute both to 
the electrical conductivity and to the magnetism.
Therefore, naively thinking, the spin-triplet state with equal-spin pairing is realized for the superconductivity.

Surprisingly the huge upper critical field of superconductivity $H_{\rm c2}$ were discovered in URhGe and UCoGe~\cite{Lev05,Miy08,Aok09_UCoGe}. 
These experimental results also support the spin triplet state, because $H_{\rm c2}$
is not limited by the Pauli paramagnetic effect which affects $H_{\rm c2}$ in the spin singlet state, instead only the orbital effect governs $H_{\rm c2}$ in the case of spin triplet state.

The superconducting properties in the ferromagnets are very sensitive to the sample qualities. 
In order to study ferromagnetic superconductivity in details, 
the high quality single crystals, fine tuning of field directions are essential. 
In this review paper, we show our recent results on ferromagnetic superconductors using our best samples~\cite{Aok11_CR,Aok11_ICHE,Aok12_JPSJ_review,Aok13_CR}. 
The superconductivity is closely related to the ferromagnetic quantum criticality.
We also show the ferromagnetic quantum critical endpoint with a fine tuning of pressure, field and temperature in UCoAl and UGe$_2$~\cite{Aok11_UCoAl,Tau10,Kot11}.
It corresponds to a collapse of the ferromagnetic ``wing'' in the temperature-pressure-field phase diagram at high fields at $P>P_{\rm c}$.

\section{Experimental}
High quality single crystals of UGe$_2$, URhGe, UCoGe and UCoAl were grown in a tetra-arc furnace using Czochralski method. 
The single crystal ingots were oriented by the Laue photograph and were cut in a spark cutter.
The ingots were subsequently annealed under ultra high vacuum. 
The quality of all single crystals was checked by the resistivity measurements using a homemade adiabatic demagnetization refrigerator (ADR) cell combined with a commercial PPMS at temperature down to $100\,{\rm mK}$.
Thanks to this simple ADR cell, we are able to check the qualities of many samples very rapidly
down to $100\,{\rm mK}$ within two hours from room temperature.
Pressure studies shown in this paper were performed using a piston cylinder or an indenter cell.
The magnetic field was applied up to $16\,{\rm T}$ and $35\,{\rm T}$ in the superconducting magnet and the resistive magnet, respectively.
The low temperature was achieved by a conventional dilution fridge and a top-loading dilution fridge.

\section{Results and Discussion}
\subsection{Crystal structure}
\begin{figure}[tbh]
\begin{center}
\includegraphics[width=0.8 \hsize,clip]{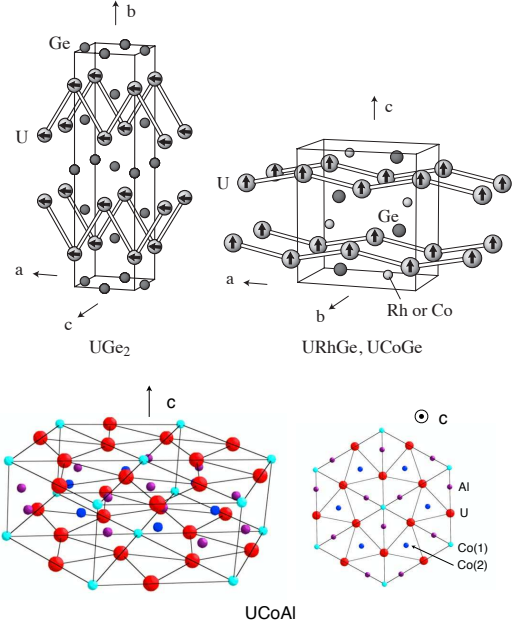}
\end{center}
\caption{(Color online) Crystal structures of UGe$_2$, URhGe, UCoGe and UCoAl.}
\label{fig:structure}
\end{figure}

Figure~\ref{fig:structure} shows the crystal structures of UGe$_2$, URhGe, UCoGe and UCoAl.
UGe$_2$ crystallizes in the orthorhombic structure with the space group $Cmmm$.
The uranium atom forms the zigzag chain with the distance $3.85\,{\rm \AA}$, which is similar 
to $\alpha$-U.
Thus the origin of $T_{\rm x}$ shown later was theoretically proposed as a trace of CDW/SDW ordering~\cite{Wat02},
however no experimental evidence was found up to now.
The magnetic moment with $1.5\,\mu_{\rm B}$ is directed along the $a$-axis.
URhGe and UCoGe belong to the same family with the TiNiSi-type orthorhombic structure (space group: $Pnma$)
The uranium atom again forms the zigzag chain along $a$-axis.
The distance is about $3.5\,{\rm \AA}$ which is close to the so-called Hill limit.
The magnetic moment with $0.4\,\mu_{\rm B}$ for URhGe and $0.05\,\mu_{\rm B}$ for UCoGe is directed along $c$-axis.
Interestingly, the magnetic moment could be slightly canted along $a$-axis due to the zigzag chain and {\it local} no inversion symmetry. 
Via an analogous Dzyaloshinski-Moriya interaction, the weak parasitic antiferromagnetism is theoretically predicted~\cite{Min06}, however, 
only the collinear ferromagnetism is experimentally found so far.
The crystal structure of UCoAl with the hexagonal ZrNiAl-type (space group: $P$\={6}2$m$) is also shown in Fig.~\ref{fig:structure}.
The uranium atom forms the quasi-kagom\'{e} lattice, indicating the possible magnetic frustration.
An interesting point is that there is no inversion symmetry in the crystal structure with the space group $P$\={6}2$m$. 

\subsection{Ferromagnetic quantum criticality}
Applying pressure in UGe$_2$, the ferromagnetism ($T_{\rm Curie}=52\,{\rm K}$ at ambient pressure)  is suppressed and the paramagnetic ground state appears.
The second order ferromagnetic transition at $T_{\rm Curie}$ changes into the first order at the tricritical point (TCP).
As shown in Fig.~\ref{fig:TPH}(a), when the field is applied in the paramagnetic state, 
UGe$_2$ shows the metamagnetic transition with the first order from the paramagnetic state to the  ferromagnetic state (FM1). 
At higher temperatures, the first order transition changes into the crossover at the critical endpoint.
The critical endpoint starting from TCP can be tuned to be $0\,{\rm K}$ which is so called quantum critical endpoint (QCEP).
The wing-shaped temperature-pressure-field phase diagram with the first order plane can be drawn.
In UGe$_2$, the QCEP is located at very high pressure ($\sim 3.5\,{\rm GPa}$) and at very high field ($\sim 20\,{\rm T}$)

On the other hand, UCoAl has already a paramagnetic ground state at ambient pressure, but is close to the ferromagnetic order.
Applying the magnetic field along $c$-axis (easy-magnetization axis) at low temperature, 
the sharp metamagnetic transition with the first order occurs at $H_{\rm m}\sim 0.6\,{\rm T}$ from the paramagnetic state to the ferromagnetic state.
The first order changes into the crossover at higher temperature, and the critical temperature $T_{\rm CEP}$ is about $10\,{\rm K}$.
With increasing pressure, $H_{\rm m}$ shifts to higher field, and $T_{\rm CEP}$ decreases and finally becomes $0\,{\rm K}$.
The QCEP is located at $H\sim 7\,{\rm T}$ and $P\sim 1.5\,{\rm GPa}$.
Further applying pressure, $H_{\rm m}$ increases further, 
but via a crossover regime, instead of the first order transition.
\begin{fullfigure}[tbh]
\begin{center}
\includegraphics[width=0.8 \hsize,clip]{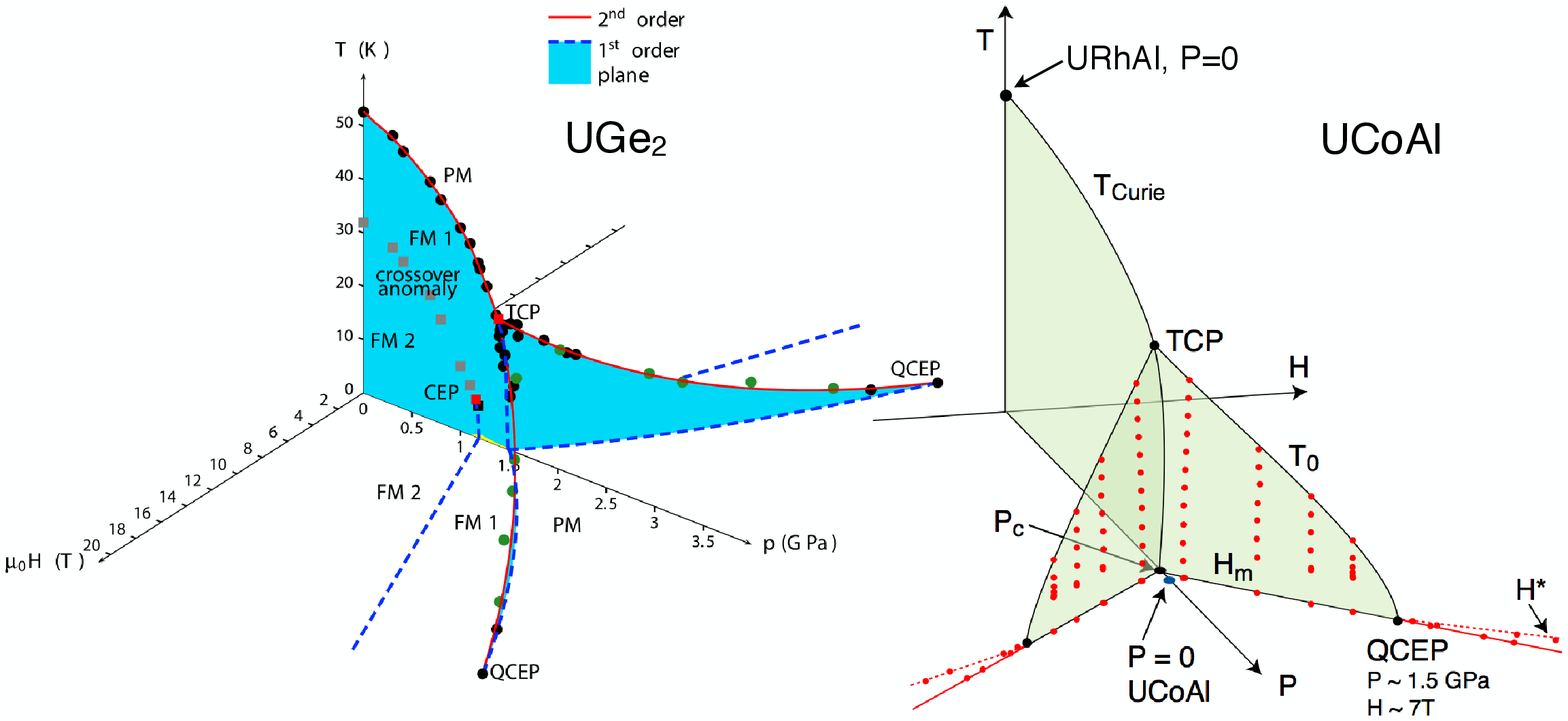}
\end{center}
\caption{(Color online) Temperature-Pressure-Field phase diagrams of UGe$_2$ and UCoAl.~\cite{Tau10,Kot11,Aok11_UCoAl}}
\label{fig:TPH}
\end{fullfigure}

\begin{figure}[tbh]
\begin{center}
\includegraphics[width=0.8 \hsize,clip]{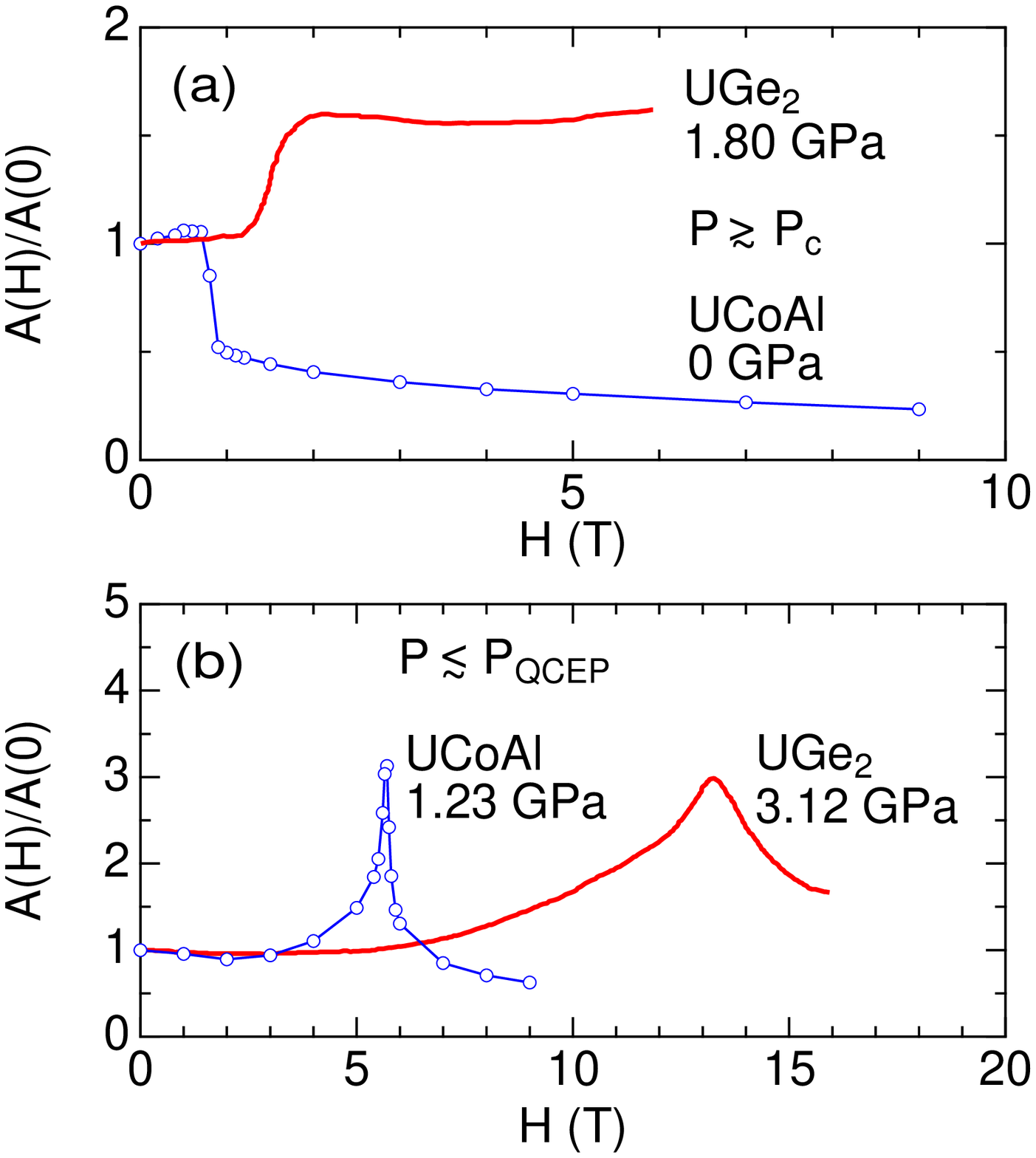}
\end{center}
\caption{(Color online) Field dependence of the resistivity $A$ coefficient in UGe$_2$ and UCoAl at low pressure (a) and at high pressure near QCEP (b).~\cite{Aok11_UCoAl}}
\label{fig:Acoef}
\end{figure}

When the ground state switches from the paramagnetic state to the ferromagnetic state at $H_{\rm m}$ with the first order,
the effective mass of conduction electrons shows the step-like behavior as a function of field, 
as shown in Fig.~\ref{fig:Acoef}(a)~\cite{Aok11_UCoAl}.
Here we assume the Kadowaki-Woods relation, namely the coefficient of $T^2$ term in resistivity, $A$ is proportional to the square of the Sommerfeld coefficient, $\gamma$.
The drastic change of the effective mass should be associated with the reconstruction of Fermi surfaces,
or with a drastic collapse of the spin fluctuations.
The $A$ coefficient in UGe$_2$ increases at $H_{\rm m}$, while in UCoAl the $A$ coefficient decreases.

On the other hand, when the pressure is tuned near QCEP,
the effective mass both in UGe$_2$ and in UCoAl shows the sharp peak at $H_{\rm m}$ as shown in Fig.~\ref{fig:Acoef}(b),
suggesting the strong magnetic fluctuations at QCEP.
Further applying pressure in UCoAl, the sharp enhancement of $A$ is smeared out, showing the broad maximum.
The Fermi surface instabilities near $H_{\rm m}$ in UCoAl were also studied by means of the thermoelectric power and Hall effect measurements.~\cite{Pal13,Com13}

\subsection{Ferromagnetism and superconductivity}
Figure~\ref{fig:TP} shows the temperature-pressure phase diagram of UGe$_2$ and URhGe and UCoGe.
$T_{\rm Curie}$ in UGe$_2$ is suppressed at $P_{\rm c}\sim 1.5\,{\rm GPa}$.
In the ferromagnetic state, there are two different ferromagnetic states named FM1 and FM2,
which are separated by $T_{\rm x}$. 
At low pressure, $T_{\rm x}$ is a crossover, but at high pressure $T_{\rm x}$ becomes the first order.
FM1 and FM2 are characterized by the different magnitude of ordered moment, $1.0\,\mu_{\rm B}$ and $1.5\,\mu_{\rm B}$, respectively.
The ordered moment suddenly changes from $1.5\,\mu_{\rm B}$ to $1.0\,\mu_{\rm B}$ at $P_{\rm x}$ when the system goes from FM2 to FM1 by applying pressure.
The Fermi surface reconstruction is associated with the transition between FM2 and FM1, and also between FM1 and the paramagnetic state.~\cite{Ter01,Set02}
The superconductivity appears only in the ferromagnetic state, namely in the pressure range of $P_{\rm x} \lesssim < P_{\rm c}$,
thus the superconductivity coexists with the ferromagnetism. 
The evidence for the microscopic coexistence, which is at least clear between $P_{\rm x}$ and $P_{\rm c}$, was given by the NQR and neutron experiments
~\cite{Kot05,Hux01}.

The superconductivity of URhGe already appears at ambient pressure at $T_{\rm sc}=0.25\,{\rm K}$, 
while $T_{\rm Curie}$ ($=9.5\,{\rm K}$) is much higher than $T_{\rm sc}$.
With pressure, $T_{\rm Curie}$ increases linearly, but $T_{\rm sc}$ decreases, indicating that
the system goes far from the critical region under pressure.

In UCoGe, the superconductivity appears again at ambient pressure at $T_{\rm sc}\sim 0.7\,{\rm K}$,
while $T_{\rm Curie}$ is about $3\,{\rm K}$.
Interestingly, $T_{\rm Curie}$ is suppressed at $P_{\rm c}\sim 1\,{\rm GPa}$,
and $T_{\rm sc}$ has a broad maximum around $P_{\rm c}$.
The superconducting phase survives even in the paramagnetic phase,
which is contradictory to the theoretical prediction by Fay and Appel,~\cite{Fay80}
where the first order nature of $T_{\rm Curie}$ is neglected.
In UCoGe, the ferromagnetic transition is of first order, inferred from the sudden jump of NQR spectra.~\cite{Oht10}
The possibility of phase separation between the paramagnetic state and the ferromagnetic state at $P=0$ 
cannot be excluded.
\begin{fullfigure}[tbh]
\begin{center}
\includegraphics[width=1 \hsize,clip]{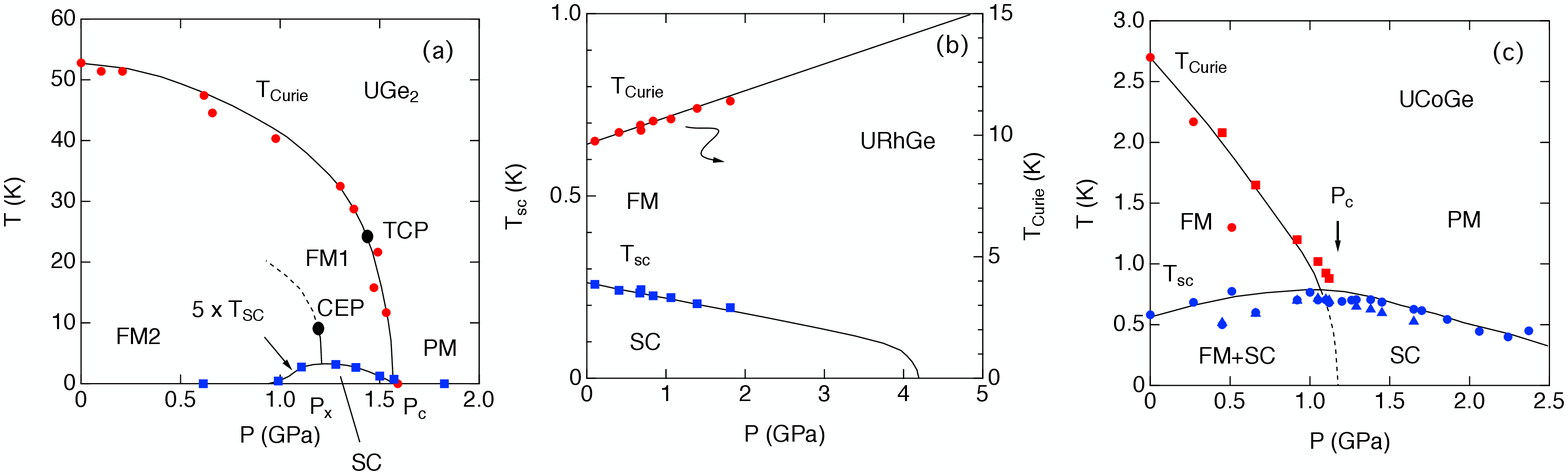}
\end{center}
\caption{(Color online) Temperature-pressure phase diagram of UGe$_2$, URhGe and UCoGe.~\cite{Sax00,Har05_pressure,Miy09,Has08_UCoGe,Slo09,Aok13_CR}}
\label{fig:TP}
\end{fullfigure}

\begin{figure}[tbh]
\begin{center}
\includegraphics[width=0.8 \hsize,clip]{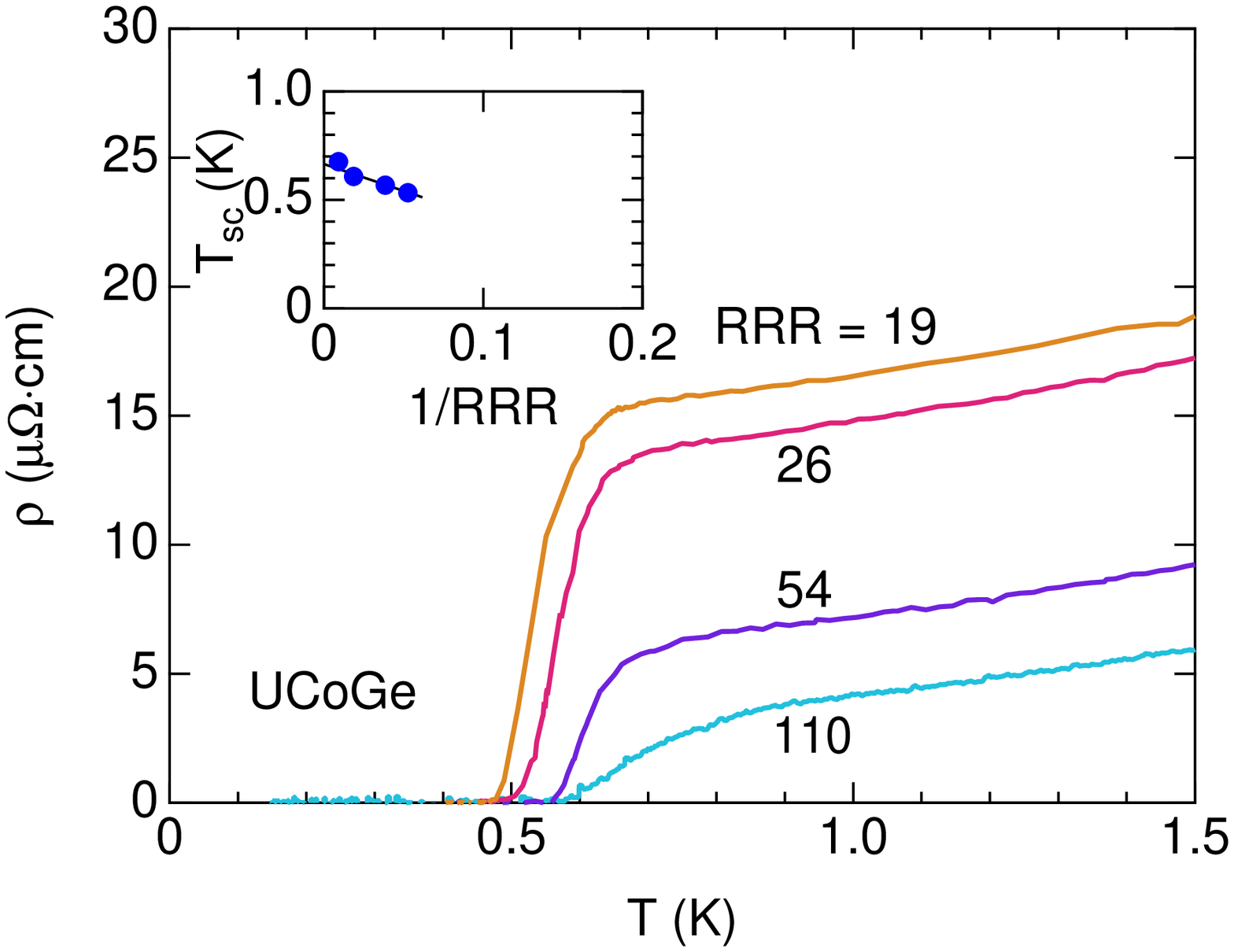}
\end{center}
\caption{(Color online) Resistivity at low temperatures in UCoGe with different quality samples. The inset shows $T_{\rm sc}$ as a function of the inverse of residual resistivity ratio (RRR).}
\label{fig:RRR}
\end{figure}

The high quality single crystals are inevitably required for the study of ferromagnetic superconductivity. 
UGe$_2$ is a rather easy material, because it has a congruent melting point.
On the other hand, URhGe and UCoGe are quite difficult to obtain the high quality,
because they are not congruent melting materials.
Therefore many attempts for the single crystal growth have been done,
by changing the composition slightly for the starting materials and the annealing conditions.
Our best samples up to now show the high residual resistivity ratio (RRR $> 100$) for URhGe and UCoGe.
The high quality was also demonstrated by the quantum oscillation measurements which is shown later.
Figure~\ref{fig:RRR} shows the resistivity data at low temperature in UCoGe for different quality samples,
and the corresponding $T_{\rm sc}$ as a function of $1/$RRR.
$T_{\rm sc}$ decreases linearly with $1/$RRR, indicating the superconductivity is affected by the sample quality.
The similar behavior is also known in URhGe, in which $T_{\rm sc}$ is more sensitive to the sample quality.~\cite{Aok03}
Basically $T_{\rm sc}$ should follow the pair-breaking theory by Abrikosov and Gor'kov~\cite{Abr61}.

Using our best quality samples of URhGe and UCoGe,
we measured the specific heat at low temperatures, as shown in Fig.~\ref{fig:Cp}~\cite{Aok13_CR}.
The data in UGe$_2$ is also shown for comparison~\cite{Tat01}.
Although the samples are in high quality, the residual $\gamma$-value in specific heat at $0\,{\rm K}$
is rather large.
Since the ordered moments of three materials are different each other, 
the residual $\gamma$-value, $\gamma_0$ was plotted as a function of the ordered moments $M_0$, as shown in Fig.~\ref{fig:Cp}.
The residual $\gamma$-value increases with $M_0$, indicating the clear correlation.
In ferromagnetic superconductors, a large internal field is created by the ordered moment. 
For example, the internal field $H_{\rm int}$ estimated from the ordered moment is $0.28\,{\rm T}$ for UGe$_2$,
$0.08\,{\rm T}$ for URhGe, and $0.01\,{\rm T}$ for UCoGe,
indicating that the system might be always in the superconducting mixed state even at zero field,
as the lower critical field $H_{\rm c1}$ is far lower than $H_{\rm int}$.
In fact, the NQR and low-temperature magnetization measurements suggest the self-induced vortex state
in UCoGe~\cite{Oht10,Deg10,Pau12_UCoGe}.
However, still no direct observation of vortex lattice has been reported.
\begin{fullfigure}[tbh]
\begin{center}
\includegraphics[width=1 \hsize,clip]{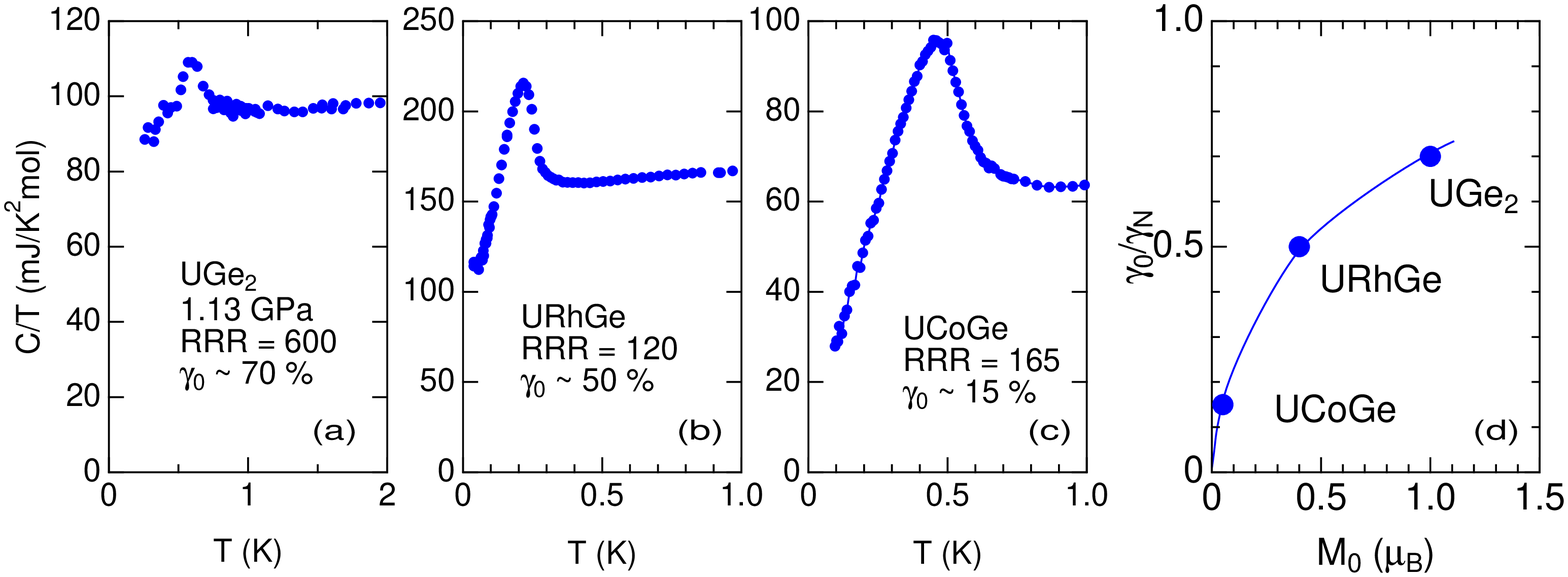}
\end{center}
\caption{(Color online) Specific heat at low temperatures in (a) UGe$_2$, (b) URhGe and (c) UCoGe. (d) Scaled residual $\gamma$-value as a function of ordered moment $M_0$.~\cite{Aok13_CR,Tat01} }
\label{fig:Cp}
\end{fullfigure}

One of the spectacular phenomena in ferromagnetic superconductors is the very large $H_{\rm c2}$.
Figure~\ref{fig:Hc2} shows the superconducting phase in the field-temperature phase diagram~\cite{She01,Lev05,Miy08,Aok09_UCoGe,Aok13_CR}.
In UGe$_2$, the field is applied along the easy magnetization axis ($a$-axis), and
the pressure is tuned just above $P_{\rm x}$.
With increasing field, the ground state is switched from FM1 to FM2.
The unusual S-shaped $H_{\rm c2}$ is observed due to this switching.
The change of Fermi surface as well as the effective mass enhancement is associated to this $H_{\rm c2}$ curve.
The value of $H_{\rm c2}$ at $0\,{\rm K}$ exceeds the Pauli limiting field expected from $T_{\rm sc}$ at zero field on the basis of 
weak coupling scheme with $g=2$,
suggesting the spin triplet state.

In URhGe, the $H_{\rm c2}$ curve is more spectacular. 
When the field is applied along the hard magnetization axis ($b$-axis),
the field-reentrant superconductivity is observed at high field range between $8\,{\rm T}$ and $13\,{\rm T}$.
The reentrant superconducting phase shows even higher $T_{\rm sc}$ ($\sim 0.4\,{\rm K}$) at $12\,{\rm T}$ than $T_{\rm sc}=0.25\,{\rm K}$
at zero field.
The superconductivity is indeed enhanced under magnetic field. 
This is contradictory to the usual superconducting behavior. 
The magnetization curve for $H\parallel b$-axis shows the relatively large initial slope at low field compared to that for $H\parallel c$-axis (easy-magnetization axis).
Further increasing field, the magnetization shows the step-like increase around $12\,{\rm T}$.
This behavior is understood by the canting process of the magnetic moment with field.
For $H \parallel b$-axis, the moment starts to tilt from $c$-axis to $b$-axis with increasing field,
and finally the moment is completely directed along $b$-axis.
In this configuration with canted moments, a scenario by Jaccarino-Peter effect can be excluded,
because the Jaccarino-Peter effect occurs when the total effective field is close to zero 
due to the compensation of external field by the internal field.
In URhGe, the moment is gradually tilted with field, which cannot make the compensation of external field.
Thus, the spin triplet state with equal spin pairing, which is free from the Pauli paramagnetic effect, should be considered.
$H_{\rm c2}$ is then governed only by the orbital effect.
If the orbital limiting field is enhanced under magnetic field for some reasons, $H_{\rm c2}$ could be enhanced as well.
We will discuss this point later.

The $H_{\rm c2}$ curve of UCoGe also displays the unusual behavior, when the field is applied along $b$-axis, as shown in Fig.~\ref{fig:Hc2}(c).
$H_{\rm c2}$ is strongly enhanced around $0.4\,{\rm K}$ with S-shape, and reaches around $18\,{\rm T}$.
In UCoGe, it seems that the reentrant phase and the low field phase observed in URhGe are merged, because of the higher $T_{\rm sc}$ at zero field.
The value of $H_{\rm c2}$ also highly exceeds the Pauli limiting field.
\begin{fullfigure}[tbh]
\begin{center}
\includegraphics[width=0.8 \hsize,clip]{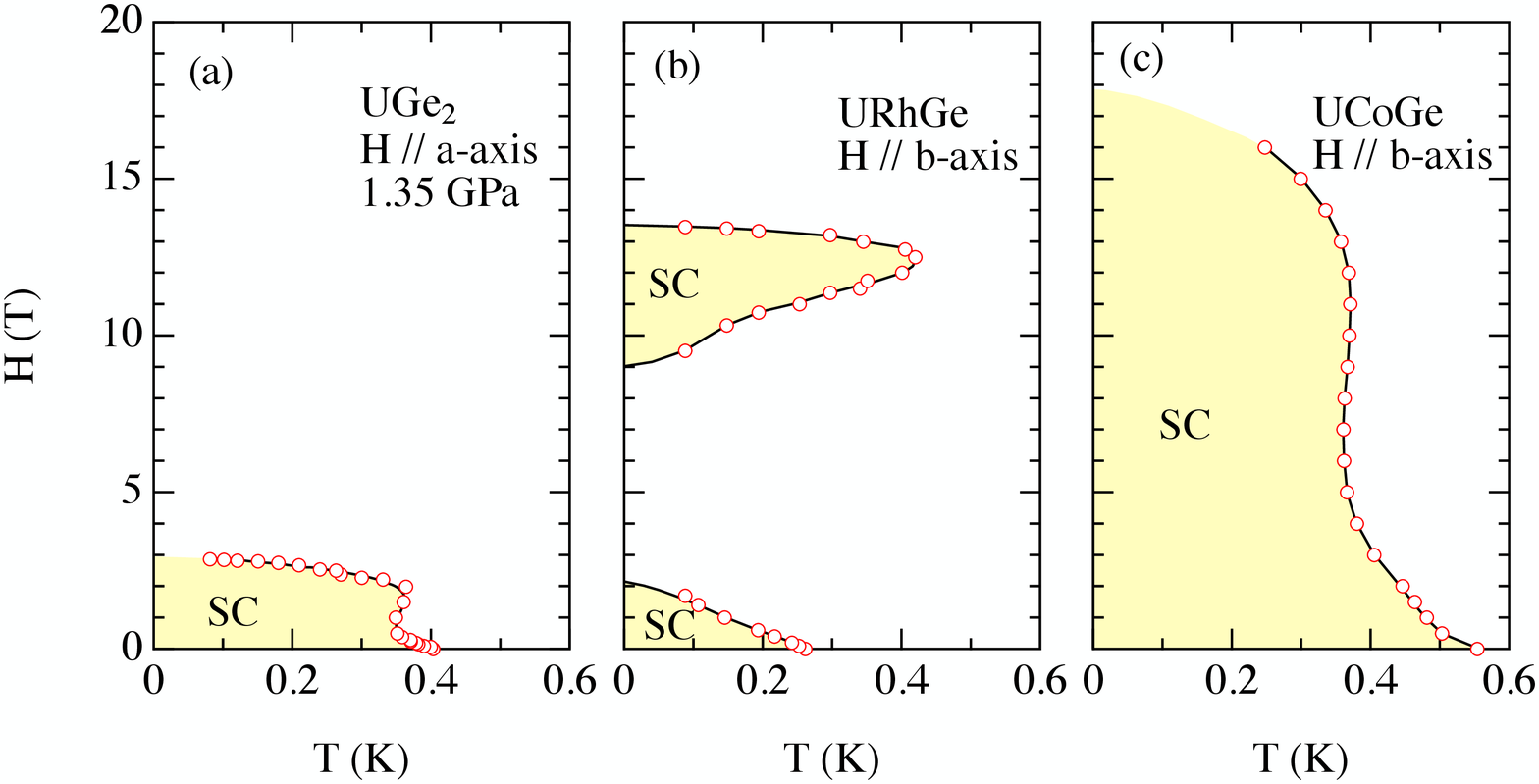}
\end{center}
\caption{(Color online) Field-temperature phase diagram at low temperatures in UGe$_2$, URhGe and UCoGe. The magnetic field is applied along the easy magnetization axis in UGe$_2$, but in URhGe and UCoGe the field direction is parallel to $b$-axis, corresponding to the hard magnetization axis.~\cite{She01,Lev05,Miy08,Aok09_UCoGe,Aok13_CR}}
\label{fig:Hc2}
\end{fullfigure}

\begin{figure}[tbh]
\begin{center}
\includegraphics[width=0.8 \hsize,clip]{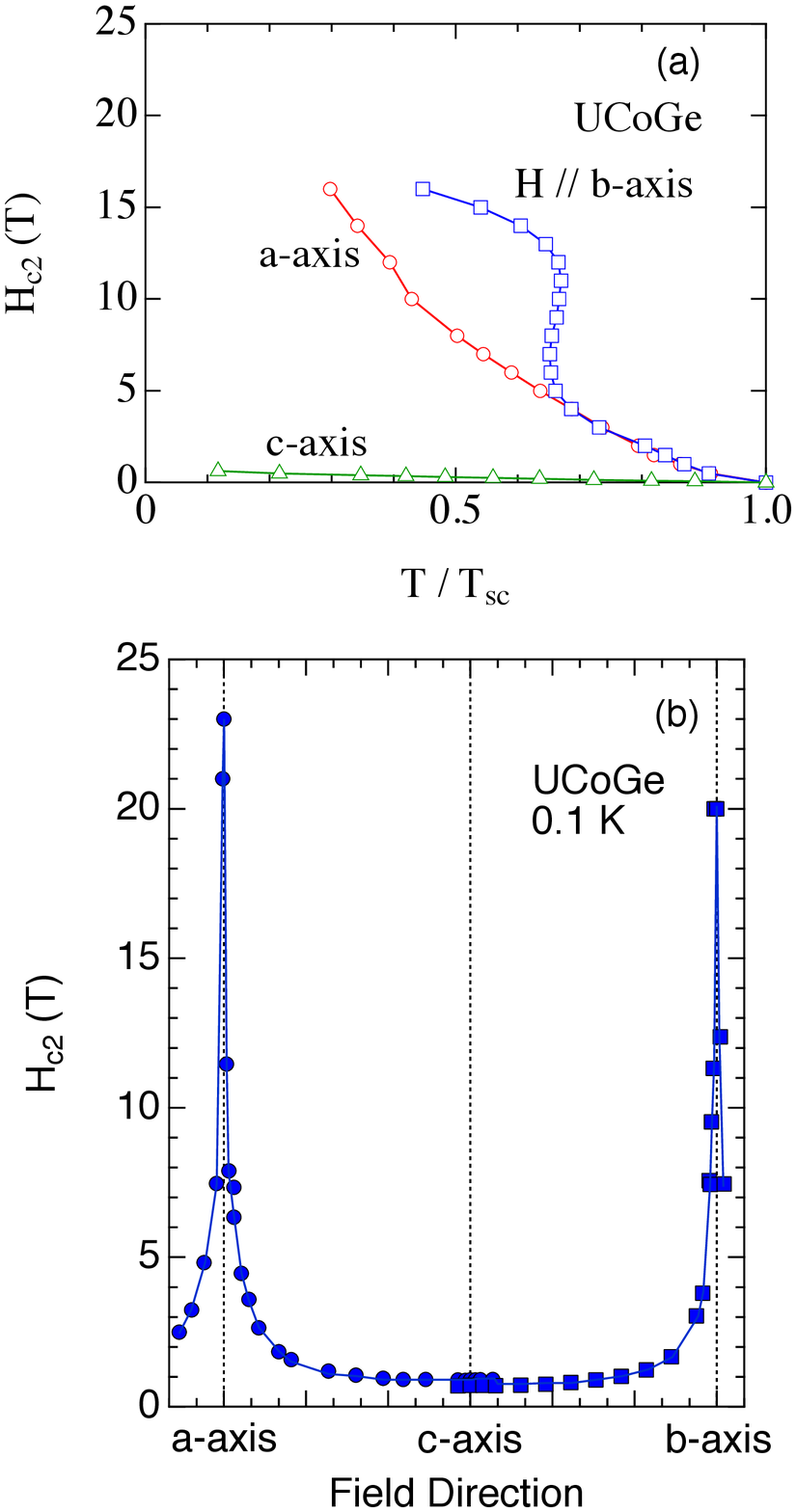}
\end{center}
\caption{(Color online) (a) Temperature dependence of $H_{\rm c2}$ for $H \parallel a$, $b$ and $c$-axis, and (b) Angular dependence of $H_{\rm c2}$ at $0.1\,{\rm K}$ in UCoGe. $T_{\rm sc}$ at $0\,{\rm K}$ is approximately $0.6\,{\rm K}$.~\cite{Aok09_UCoGe}}
\label{fig:UCoGe_Hc2}
\end{figure}

The high $H_{\rm c2}$ is very sensitive to the field direction to the sample in UCoGe and URhGe.
Figure~\ref{fig:UCoGe_Hc2}(a) shows the temperature dependence of $H_{\rm c2}$ for different field directions in UCoGe.
The angular dependence of $H_{\rm c2}$ is shown in Fig.~\ref{fig:UCoGe_Hc2}(b).
When the field is applied along $a$-axis which corresponds to the hardest magnetization axis,
$H_{\rm c2}$ at $0\,{\rm K}$ seems to be even higher than that for $H\parallel b$-axis, showing the upward curvature with decreasing temperature.
The angular dependence of $H_{\rm c2}$ at $0.1\,{\rm K}$ is shown in Fig.~\ref{fig:UCoGe_Hc2}(b).
If the field direction is slightly tilted to $c$-axis (easy-magnetization axis),
$H_{\rm c2}$ is immediately suppressed.
This angular dependence cannot be explained by the conventional effective mass model assuming 
the ellipsoidal Fermi surface associated with the anisotropic effective mass.
An alternative mechanism which explains the very anisotropic $H_{\rm c2}$ should be considered.

\begin{figure}[tbh]
\begin{center}
\includegraphics[width=1 \hsize,clip]{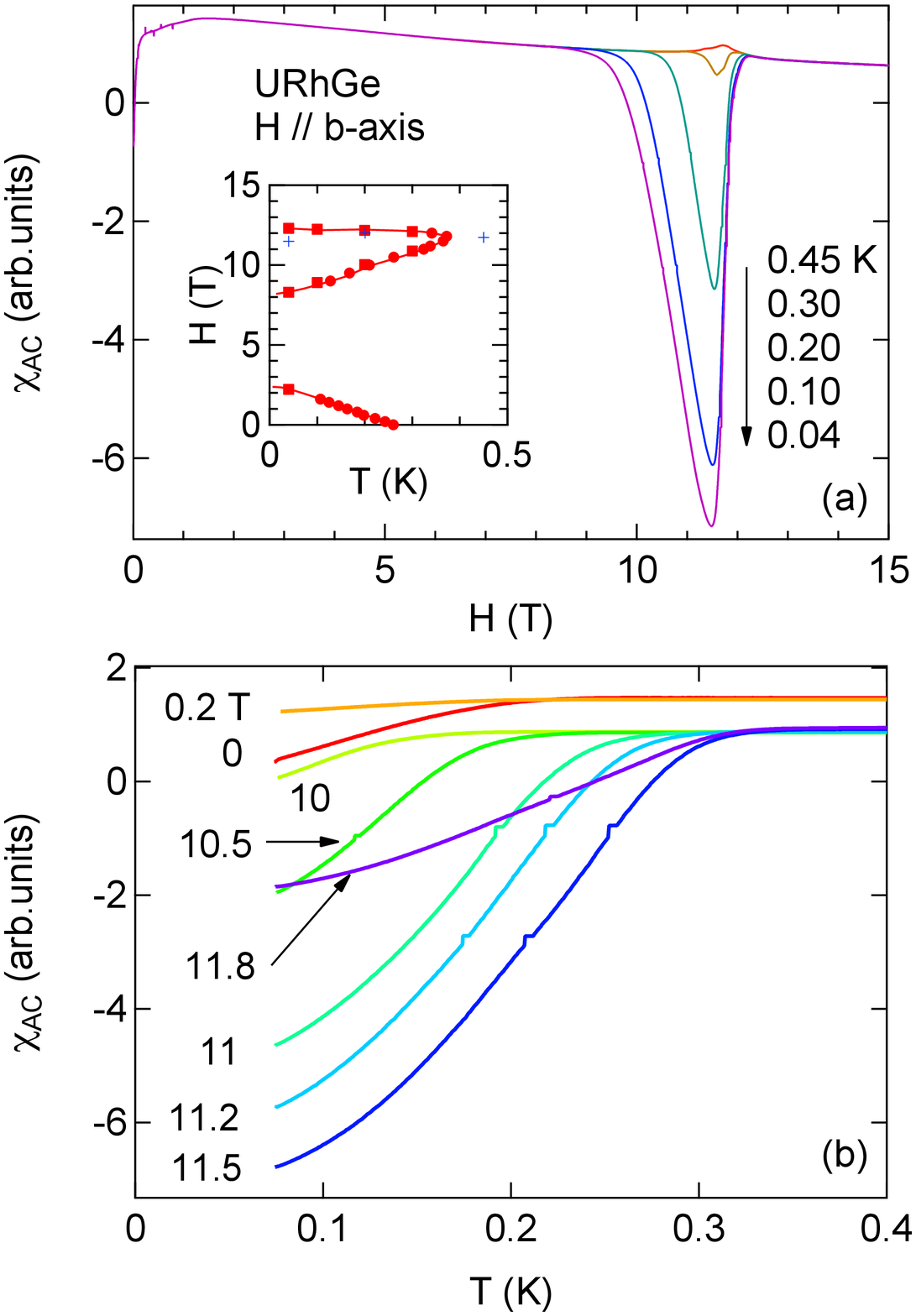}
\end{center}
\caption{(Color online) (a) Field and (b) temperature dependence of AC susceptibility for $H\parallel b$-axis in URhGe. The inset shows the field-temperature phase diagram }
\label{fig:URhGe_AC_chi}
\end{figure}

Another interesting feature is the reentrant superconductivity is very robust compare to the low-field superconductivity. 
Figure~\ref{fig:URhGe_AC_chi} shows the field and temperature dependence of AC susceptibility for $H\parallel b$-axis in URhGe.
The inset shows the field-temperature phase diagram for superconductivity defined by the onset of anomaly in the AC susceptibility measurements. 
The phase diagram by AC susceptibility measurements is in good agreement with that obtained by the 
resistivity measurements in the same sample.
Surprisingly, the drop of AC susceptibility at high fields due to the diamagnetic signal of superconductivity
is much larger than that at low fields, indicating the robust superconductivity at high fields.
It might be also associated with the unusual vortex state at high fields,
although the microscopic evidence is not obtained yet.
The similar results in the AC susceptibility are also obtained in UCoGe.

\begin{figure}[tbh]
\begin{center}
\includegraphics[width=1 \hsize,clip]{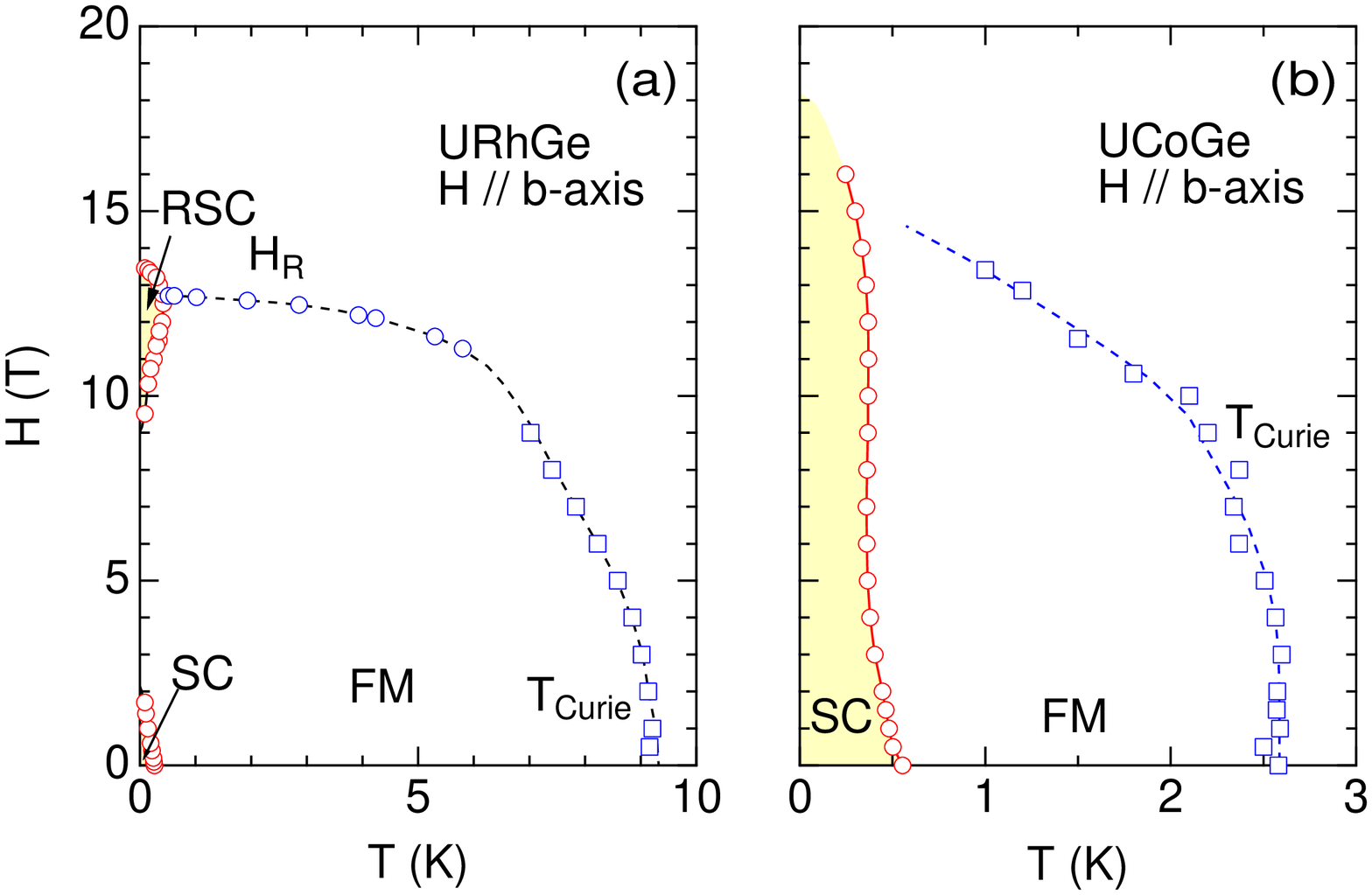}
\end{center}
\caption{(Color online) Field-temperature phase diagram of URhGe and UCoGe for the field along the hard magnetization axis ($b$-axis).~\cite{Aok09_UCoGe,Miy08,Aok11_CR}}
\label{fig:HT}
\end{figure}

A key feature as a reason for the appearance of field-reentrant (-reinforced) superconductivity is the
suppression of $T_{\rm Curie}$ at transversal high fields in terms of easy-magnetization axis.
Figure~\ref{fig:HT} shows the field-temperature phase diagram of URhGe and UCoGe
when the field is applied along the $b$-axis.
In general $T_{\rm Curie}$ at high fields is not well defined in ferromagnets,
because the phase transition immediately becomes the broad crossover between the paramagnetic state and the field-induced ferromagnetic state.
However, when the field is perfectly aligned to the hard-magnetization axis in the Ising system,
$T_{\rm Curie}$ can be clearly defined even at high fields.
$T_{\rm Curie}$ should decrease with fields, following the relation with $\Delta T_{\rm Curie}\propto -H^2$,
according to the theory~\cite{Min11}.
In fact, $T_{\rm Curie}$ of URhGe and UCoGe decrease with fields and is suppressed at $\sim 13\,{\rm T}$ 
and $\sim 15\,{\rm T}$, respectively.
The superconducting phase is connected to the suppressed $T_{\rm Curie}$ in the phase diagram both in URhGe
and in UCoGe.
One can naively believe that the enhancement of ferromagnetic fluctuations at high field play an important role for field-reentrant (-reinforced) superconductivity.

\begin{figure}[tbh]
\begin{center}
\includegraphics[width=1 \hsize,clip]{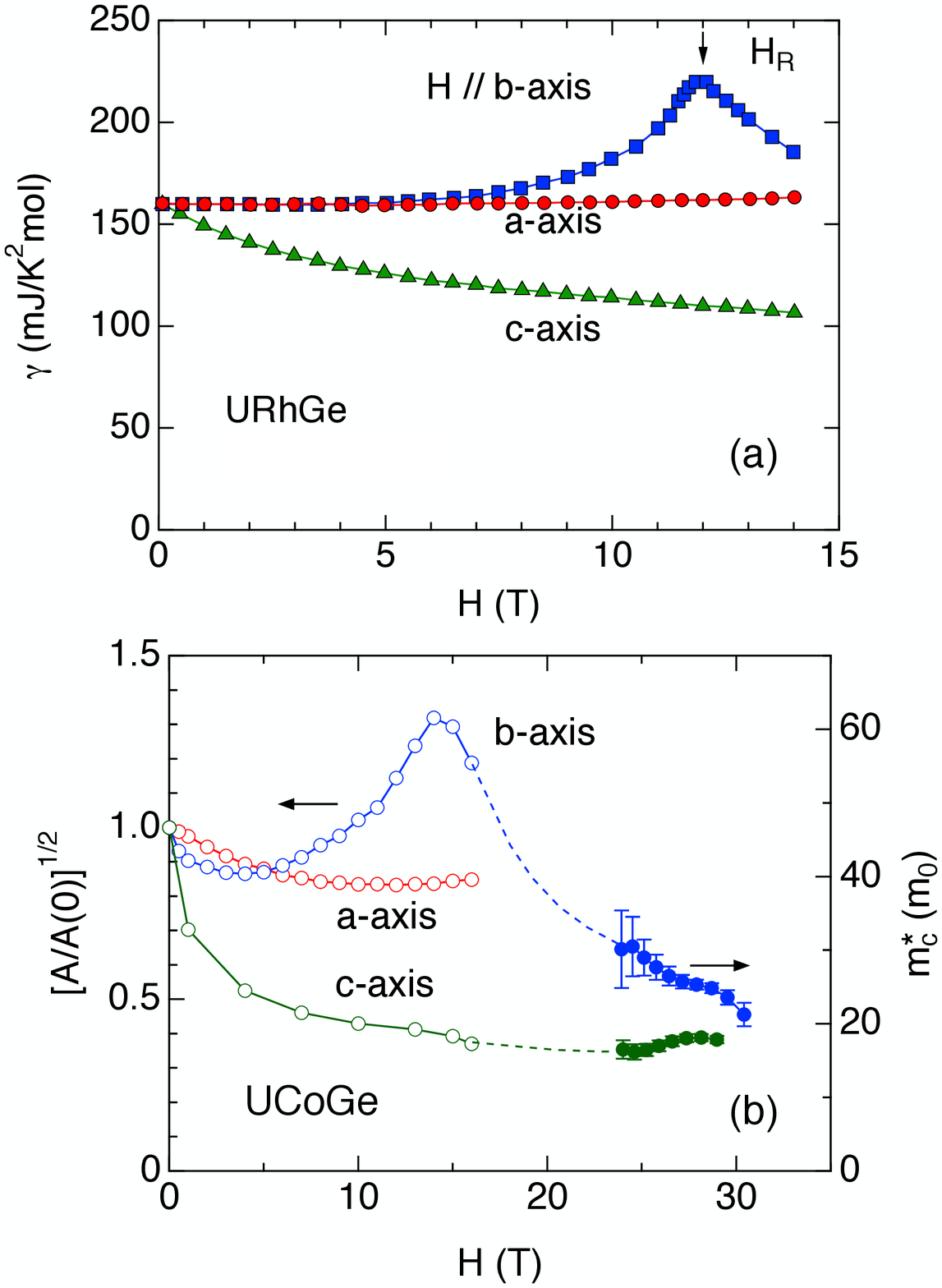}
\end{center}
\caption{(Color online) (a) Field dependence of Sommerfeld coefficient in URhGe for $H\parallel a$, $b$ and $c$-axis. (b) Field dependence of resistivity $A$ coefficient in the form of $[A(H)/A(0)]^{1/2}$ vs $H$ in UCoGe. High fields data above $20\,{\rm T}$ is obtained by the SdH experiments.~\cite{Har11,Aok09_UCoGe,Aok11_UCoGe}}
\label{fig:Hdep_mass}
\end{figure}

Figure~\ref{fig:Hdep_mass} shows the field-dependence of effective mass in URhGe and UCoGe determined by the Sommerfeld coefficient $\gamma$, the resistivity $A$ coefficient, assuming the validity of Kadowaki-Woods ratio, and the Shubnikov-de Haas experiments.
The effective mass is clearly enhanced at high fields,
when the field is applied along $b$-axis.
On the other hand, the effective mass decreases monotonously with field for the field along the easy-magnetizaton axis ($c$-axis).
These results suggest that the effective mass is enhanced when the ferromagnetic fluctuation is induced at transversal high fields,
whereas the ferromagnetic fluctuation is reduced in the longitudinal configuration, which has been indeed observed in NMR experiments~\cite{Hat12}.

In general, $H_{\rm c2}$ is governed by two effects, namely the Pauli paramagnetic effect and the orbital effect.
In ferromagnetic superconductors, there is no Pauli paramagnetic effect because of the spin-triplet state with equal spin paring. 
Therefore the $H_{\rm c2}$ is limited only by the orbital effect.
The orbital limiting field $H_{\rm orb}$ is described by the coherence length $\xi$, namely $H_{\rm orb}\propto 1/\xi^2$.
The coherence length $\xi$ can be described by $\xi \sim \hbar v_{\rm F}/(k_{\rm B}T_{\rm sc})$,
and the Fermi velocity $v_{\rm F}$ has a relation of $m^\ast v_{\rm F} = \hbar k_{\rm F}$,
where $k_{\rm F}$ is the Fermi wave number.
Thus $H_{\rm orb}$ is simply described by $H_{\rm orb}\sim (m^\ast T_{\rm sc})^2$.
Furthermore, $T_{\rm sc}$ is also written by $T_{\rm sc}\sim \exp [-(\lambda +1)/\lambda]$,
where $\lambda$ has a relation between $m^\ast$ and the band mass $m_{\rm b}$, namely $m^\ast = (1+\lambda) m_{\rm b}= m_{\rm b} + m^{\ast\ast}$. 
If the effective mass is enhanced due to the enhancement of the ``correlation'' mass $m^{\ast\ast}$ which is linked to the ferromagnetic fluctuations, $T_{\rm sc}$ is also enhanced.
Thus $H_{\rm orb}$ is further increased.
In this crude model, we can explain the field-reentrant (-reinforced) superconductivity in URhGe and UCoGe.

\begin{figure}[tbh]
\begin{center}
\includegraphics[width=1 \hsize,clip]{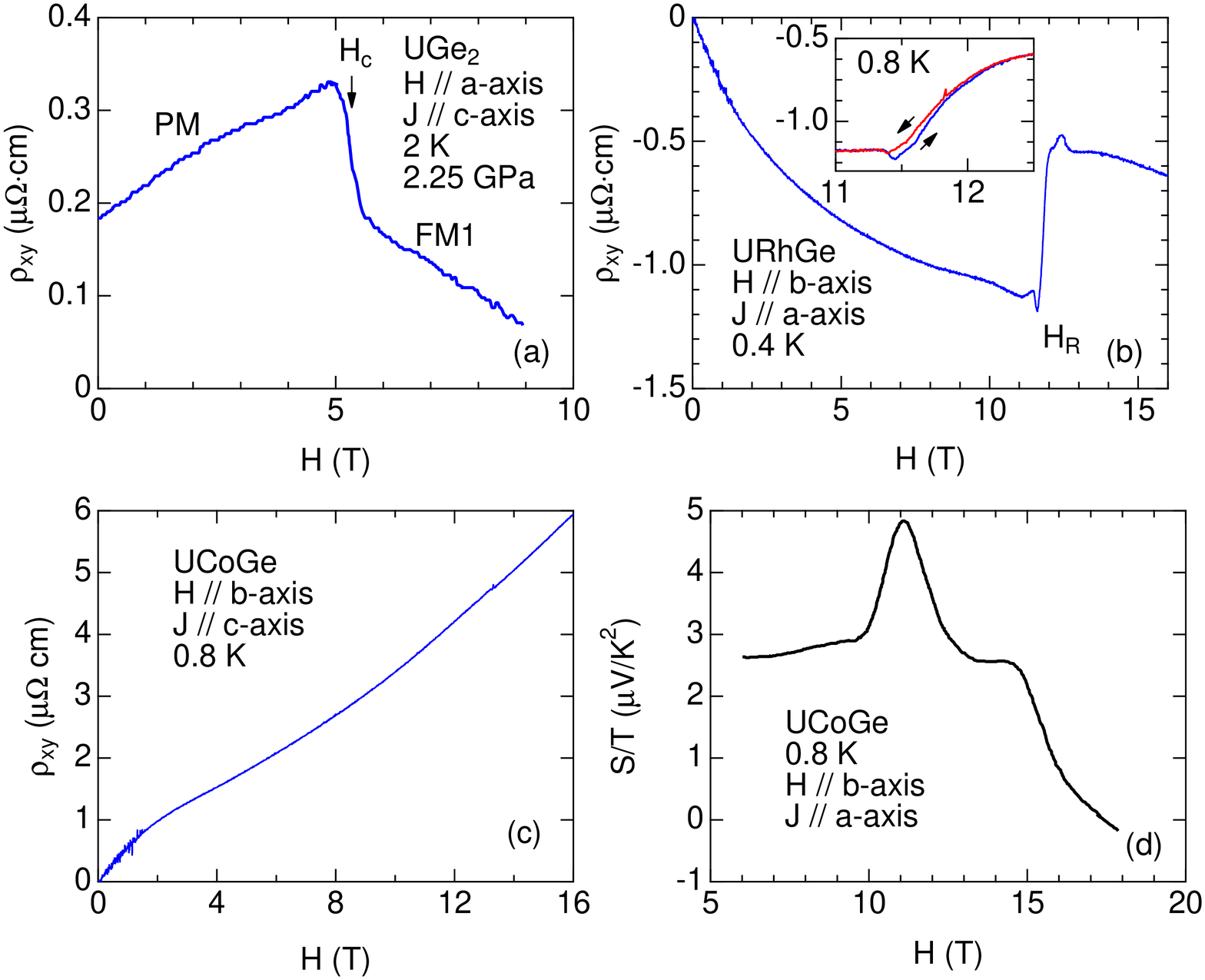}
\end{center}
\caption{(Color online) Field dependence of Hall resistivity in (a) UGe$_2$, (b) URhGe and (c) UCoGe at low temperatures. The inset in panel (b) shows the hysteresis of Hall resistivity at $0.8\,{\rm K}$ which is far above $T_{\rm sc}$ for reentrant superconductivity. (d) Field dependence of thermoelectric power in UCoGe.~\cite{Kot11,Mal12}}
\label{fig:Hall}
\end{figure}

Here we assumed that the Fermi surface is unchanged under magnetic fields. 
In reality, a drastic change of Fermi surface can be expected in URhGe and UCoGe, 
when $T_{\rm Curie}$ is suppressed at high fields.
Figure~\ref{fig:Hall} shows the field dependence of Hall resistivity at low temperatures in URhGe and UCoGe together with UGe$_2$.
Although the interpretation for the field-response of Hall resistivity is quite difficult because of the anomalous Hall effect,
the sudden jumps in UGe$_2$ and URhGe imply the Fermi surface reconstruction at high fields.
The field dependence of Hall resistivity in UCoGe shows no anomaly up to $16\,{\rm T}$,
but thermoelectric power again shows the anomaly around $12\,{\rm T}$,
suggesting the change of Fermi surface.

Considering the change of Fermi surface, the orbital limit should be rewritten by $H_{\rm orb}\sim (m^\ast T_{\rm sc}/k_{\rm F})^2$.
One can expect that $H_{\rm orb}$ is increased when the $k_{\rm F}$ is suppressed,
as it may happen due to the Lifshitz transition for one specific band.
In fact, UCoGe shows the volume change of Fermi surface with heavy effective mass at high fields in the SdH experiments~\cite{Aok11_UCoGe}.
In URhGe the similar volume change of the pocket Fermi surface is reported~\cite{Yel11}.
However, the consequence of the disappearance of one single orbit on superconductivity in the multi-band system is not a trivial problem,
and thus will deserve careful theoretical configurations.

\begin{figure}[tbh]
\begin{center}
\includegraphics[width=0.8 \hsize,clip]{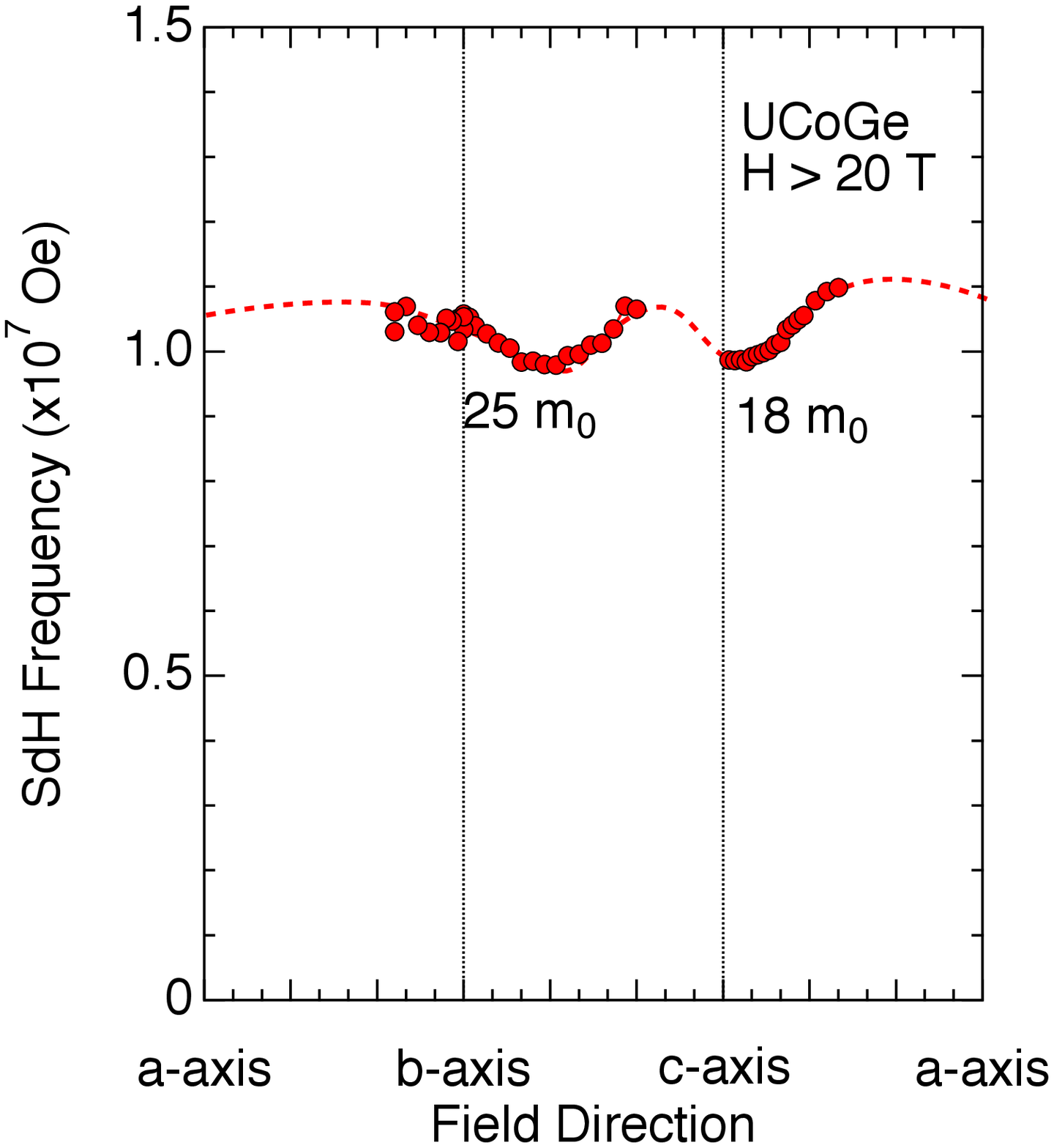}
\end{center}
\caption{(Color online) Angular dependence of SdH frequency above $20\,{\rm T}$ in UCoGe. The detected cyclotron effective mass is also shown for $H \parallel b$ and $c$-axis.~\cite{Aok11_UCoGe}}
\label{fig:dHvA}
\end{figure}

Finally we show the angular dependence of SdH frequency obtained at high fields above $20\,{\rm T}$ in UCoGe~. 
Since the sample quality is still not sufficient for quantum oscillation measurements,
the detected Fermi surface is only one in the limited field direction.
Figure~\ref{fig:dHvA} shows the angular dependence of SdH frequency above $20\,{\rm T}$ at low temperatures in UCoGe.
The SdH frequency $F$ is proportional to the cross-sectional area of Fermi surface $S_{\rm F}$,
namely $F=\hbar c S_{\rm F}/(2\pi e)$.
The frequency does not show the significant angular dependence in the detected field angle range,
indicating the small pocket Fermi surface with spherical shape.
The volume of Fermi surface occupies $2\,\%$ in the Brillouin zone,
while the corresponding $\gamma$-value is $7\,{\rm mJ\,K^{-2}mol^{-1}}$.
Although the Fermi surface is small in volume,
the contribution to the total $\gamma$-value ($55\,{\rm mJ\,K^{-2}mol^{-1}}$ at zero field)
reaches $13\,\%$
This implies that UCoGe is a low carrier system associated with heavy quasi-particles.
In fact, the band structure calculation based on the 5$f$-itinerant model shows the relatively small Fermi surface.~\cite{Sam10}
The ratio of thermoelectric power and Sommerfeld coefficient, so-called $q$-factor, ($q = (S/T) N_{\rm A} e/\gamma$) is $5$ in UCoGe,~\cite{Mal12}
which is comparable to the low carrier heavy fermion compound URu$_2$Si$_2$.

In these low carrier heavy fermion system, a fascinating field-effect is expected.
The effective Fermi energy of Fermi surface can be written as $\varepsilon_{\rm F}= \hbar^2 k_{\rm F}^2/(2m^\ast)$.
If the Fermi surface is small in volume and the effective mass is large, 
the effective Fermi energy becomes small,
which could be comparable to the Zeeman energy induced by the magnetic field.
In URu$_2$Si$_2$, the change of Fermi surface occurs at high fields, depending on the 
carrier density and effective mass.
The Fermi surface of UCoGe also shows the field-dependent SdH frequency and cyclotron mass, which is
 observed above $20\,{\rm T}$.
In Fig.~\ref{fig:Hdep_mass}(b), the decrease of cyclotron mass for $H\parallel b$-axis from $30\,m_0$ to $20\,m_0$ with field is shown.

\section{Summary}
We reviewed our studies on ferromagnetic superconductors, UGe$_2$, URhGe and UCoGe.
High quality single crystals and precise tuning for field direction and pressure 
are very important in order to investigate the peculiar superconducting properties.
All ferromagnetic superconductors show the large $H_{\rm c2}$, which highly exceeds 
the Pauli paramagnetic limit,
indicating the formation of spin-triplet state with equal-spin pairing.
The field-reentrant (-reinforced) superconductivity can be interpreted by the enhancement of effective mass, which is coupled to the ferromagnetic fluctuations.
Associated Fermi surface instabilities, such as Lifshitz-like transition may also play a key role.
The ferromagnetic quantum critical endpoint in the temperature-field-pressure phase diagram 
was clarified in UGe$_2$ with clear interplay between ferromagnetic enhancement and Fermi surface reconstructions.
To elucidate this interplay, the determination of Fermi surface is a key issue in UCoAl.
The sharp peak of resistivity $A$ coefficient at $H_{\rm m}$ as a function of field
indicates that the ferromagnetic fluctuations are enhanced at quantum critical endpoint.

\section*{Acknowledgements}
We acknowledge
J. P. Brison, T. Combier, S. Fujimoto, F. Hardy, H. Harima, K. Hasselbach, E. Hassinger, T. Hattori, L. Howald, D. Hykel, K. Ishida, N. Kimura, W. Knafo, G. Knebel, H. Kotegawa, L. Malone, T. D. Matsuda, C. Meingast, V. Mineev, A. Miyake, K. Miyake, C. Paulsen, S. Raymond, G. Scheerer, I. Sheikin, Y. Tada, V. Taufour, M. Taupin and H. Yamagami.
This work was financially supported by ERC starting grant (NewHeavyFermion), French ANR project (CORMAT, SINUS, DELICE), REIMEI, ICC-IMR and KAKENHI.


\begin{thebibliography}{10}

\bibitem{Fer77}
W.~A. Fertig, D.~C. Johnston, L.~E. DeLong, R.~W. McCallum, M.~B. Maple and B.~T. Matthias: 
Phys. Rev. Lett. {\bf 38}, 987 (1977).

\bibitem{Ish77}
M.~Ishikawa and O.~Fischer: 
Solid State Commun. {\bf 23}, 37 (1977).

\bibitem{Sax00}
S.~S. Saxena, P.~Agarwal, K.~Ahilan, F.~M. Grosche, R.~K.~W. Haselwimmer, M.~J.
  Steiner, E.~Pugh, I.~R. Walker, S.~R. Julian, P.~Monthoux, G.~G. Lonzarich,
  A.~Huxley, I.~Sheikin, D.~Braithwaite and J.~Flouquet: 
Nature {\bf 406}, 587 (2000).

\bibitem{Aok01}
D.~Aoki, A.~Huxley, E.~Ressouche, D.~Braithwaite, J.~Flouquet, J.-P. Brison, E.~Lhotel and C.~Paulsen: 
  Nature {\bf 413}, 613 (2001).

\bibitem{Huy07}
N.~T. Huy, A.~Gasparini, D.~E. {de Nijs}, Y.~Huang, J.~C.~P. Klaasse,
  T.~Gortenmulder, A.~{de Visser}, A.~Hamann, T.~{G\"{o}rlach} and
  H.~v.~{L\"{o}hneysen}: 
  Phys. Rev. Lett. {\bf 99}, 067006  (2007).

\bibitem{Kot05}
H.~Kotegawa, A.~Harada, S.~Kawasaki, Y.~Kawasaki, Y.~Kitaoka, Y.~Haga,
  E.~Yamamoto, Y.~\={O}nuki, K.~M. Itoh, E.~E. Haller and H.~Harima: 
  J. Phys. Soc. Jpn. {\bf 74}, 705 (2005).

\bibitem{Hux03}
A.~Huxley, E.~Ressouche, B.~Grenier, D.~Aoki, J.~Flouquet and C.~Pfleiderer: 
J. Phys.: Condens. Matter {\bf 15}, S1945 (2003).

\bibitem{Oht08}
T.~Ohta, Y.~Nakai, Y.~Ihara, K.~Ishida, K.~Deguchi, N.~K. Sato and I.~Satoh: 
J. Phys. Soc. Jpn. {\bf 77}, 023707 (2008).

\bibitem{Hat13}
T.~Hattori, K.~Karube, Y.~Ihara, K.~Ishida, K.~Deguchi, N.~K. Sato and
  T.~Yamamura: 
  Phys. Rev. B {\bf 88}, 085127 (2013).


\bibitem{Vis09}
A.~{de Visser}, N.~T. Huy, A.~Gasparini, D.~E. {de Nijs}, D.~Andreica, C.~Baines and A.~Amato: 
Phys. Rev. Lett. {\bf 102}, 167003 (2009).

\bibitem{Lev05}
F.~L\'{e}vy, I.~Sheikin, B.~Grenier and A.~D. Huxley: 
Science {\bf 309}, 1343 (2005).

\bibitem{Miy08}
A.~Miyake, D.~Aoki and J.~Flouquet: 
J. Phys. Soc. Jpn. {\bf 77}, 094709 (2008).

\bibitem{Aok09_UCoGe}
D.~Aoki, T.~D. Matsuda, V.~Taufour, E.~Hassinger, G.~Knebel and J.~Flouquet: 
J. Phys. Soc. Jpn. {\bf 78}, 113709 (2009).

\bibitem{Aok11_CR}
D.~Aoki, F.~Hardy, A.~Miyake, V.~Taufour, T.~D. Matsuda and J.~Flouquet: 
C. R.  Physique {\bf 12}, 573 (2011).

\bibitem{Aok11_ICHE}
D.~Aoki, T.~D. Matsuda, F.~Hardy, C.~Meingast, V.~Taufour, E.~Hassinger,
  I.~Sheikin, C.~Paulsen, G.~Knebel, H.~Kotegawa and J.~Flouquet: 
  J. Phys. Soc. Jpn. {\bf 80}, SA008 (2011).

\bibitem{Aok12_JPSJ_review}
D.~Aoki and J.~Flouquet: 
J. Phys. Soc. Jpn. {\bf 81}, 011003 (2012).

\bibitem{Aok13_CR}
D.~Aoki, W.~Knafo and I.~Sheikin: 
C. R. Physique {\bf 14}, 53 (2013).

\bibitem{Aok11_UCoAl}
D.~Aoki, T.~Combier, V.~Taufour, T.~D. Matsuda, G.~Knebel, H.~Kotegawa and J.~Flouquet: 
J. Phys. Soc. Jpn. {\bf 80}, 094711 (2011).

\bibitem{Tau10}
V.~Taufour, D.~Aoki, G.~Knebel and J.~Flouquet: 
Phys. Rev. Lett. {\bf 105}, 217201 (2010).

\bibitem{Kot11}
H.~Kotegawa, V.~Taufour, D.~Aoki, G.~Knebel and J.~Flouquet: 
J. Phys. Soc. Jpn. {\bf 80}, 083703 (2011).

\bibitem{Wat02}
S.~Watanabe and K.~Miyake: 
J. Phys. Soc. Jpn. {\bf 71}, 2489 (2002).

\bibitem{Min06}
V.~P. Mineev: 
C. R. Physique {\bf 7}, 35 (2006).

\bibitem{Pal13}
A.~Palacio-Morales, A.~Pourret, G.~Knebel, T.~Combier, D.~Aoki, H.~Harima and J.~Flouquet: 
Phys. Rev. Lett. {\bf 110}, 116404 (2013).

\bibitem{Com13}
T.~Combier, D.~Aoki, G.~Knebel and J.~Flouquet: 
J. Phys. Soc. Jpn. {\bf 82}, 104705 (2013).

\bibitem{Ter01}
T.~Terashima, T.~Matsumoto, C.~Terakura, S.~Uji, N.~Kimura, M.~Endo, T.~Komatsubara and H.~Aoki: 
Phys. Rev. Lett. {\bf 87}, 166401 (2001).

\bibitem{Set02}
R.~Settai, M.~Nakashima, S.~Araki, Y.~Haga, T.~C. Kobayashi, N.~Tateiwa, H.~Yamagami and Y.~\={O}nuki: 
J. Phys.: Condens. Matter {\bf 14}, L29 (2002).

\bibitem{Hux01}
A.~Huxley, I.~Sheikin, E.~Ressouche, N.~Kernavanois, D.~Braithwaite, R.~Calemczuk and J.~Flouquet: 
Phys. Rev. B {\bf 63}, 144519 (2001).

\bibitem{Fay80}
D.~Fay and J.~Appel: 
Phys. Rev. B {\bf 22}, 3173 (1980).

\bibitem{Oht10}
T.~Ohta, T.~Hattori, K.~Ishida, Y.~Nakai, E.~Osaki, K.~Deguchi, N.~K. Sato and I.~Satoh: 
J. Phys. Soc. Jpn. {\bf 79}, 023707 (2010).

\bibitem{Har05_pressure}
F.~Hardy, A.~Huxley, J.~Flouquet, B.~Salce, G.~Knebel, D.~Braithwaite, D.~Aoki, M.~Uhlarz and C.~Pfleiderer: 
Physica B {\bf 359}, 1111 (2005).

\bibitem{Miy09}
A.~Miyake, D.~Aoki and J.~Flouquet:
J. Phys. Soc. Jpn. {\bf 78}, 063703 (2009).

\bibitem{Has08_UCoGe}
E.~Hassinger, D.~Aoki, G.~Knebel and J.~Flouquet: 
J. Phys. Soc. Jpn. {\bf 77}, 073703 (2008).

\bibitem{Slo09}
E.~Slooten, T.~Naka, A.~Gasparini, Y.~K. Huang and A.~de~Visser: 
Phys. Rev. Lett. {\bf 103}, 097003 (2009).

\bibitem{Aok03}
D.~Aoki, A.~Huxley, F.~Hardy, D.~Braithwaite, E.~Ressouche, J.~Flouquet, J.~P. Brison and C.~Paulsen: 
Acta Phys. Pol. B {\bf 34}, 503 (2003).

\bibitem{Abr61}
A.~A. Abrikosov and L.~P. Gor'kov: 
Sov. Phys. JETP {\bf 1243}, 12 (1961).

\bibitem{Tat01}
N.~Tateiwa, T.~C. Kobayashi, K.~Hanazono, K.~Amaya, Y.~Haga, R.~Settai and Y.~\={O}nuki: 
J. Phys.: Condens. Matter {\bf 13}, L17 (2001).

\bibitem{Deg10}
K.~Deguchi, E.~Osaki, S.~Ban, N.~Tamura, Y.~Simura, T.~Sakakibara, I.~Satoh and N.~K. Sato: 
J. Phys. Soc. Jpn. {\bf 79}, 083708 (2010).

\bibitem{Pau12_UCoGe}
C.~Paulsen, D.~J. Hykel, K.~Hasselbach and D.~Aoki: 
Phys. Rev. Lett. {\bf 109}, 237001 (2012).

\bibitem{She01}
I.~Sheikin, A.~Huxley, D.~Braithwaite, J.~P. Brison, S.~Watanabe, K.~Miyake and J.~Flouquet: 
Phys. Rev. B {\bf 64}, 220503 (2001).

\bibitem{Min11}
V.~P. Mineev: Phys. Rev. B {\bf 83}, 064515 (2011).

\bibitem{Hat12}
T.~Hattori, Y.~Ihara, Y.~Nakai, K.~Ishida, Y.~Tada, S.~Fujimoto, N.~Kawakami, E.~Osaki, K.~Deguchi, N. K.~Sato and I.~Satoh:
 Phys. Rev. Lett. {\bf 108}, 066403 (2012).

\bibitem{Har11}
F.~Hardy, D.~Aoki, C.~Meingast, P.~Schweiss, P.~Burger, H.~v.~Loehneysen and J.~Flouquet: 
Phys. Rev. B {\bf 83} 195107 (2011).

\bibitem{Aok11_UCoGe}
D.~Aoki, I.~Sheikin, T.~D. Matsuda, V.~Taufour, G.~Knebel and J.~Flouquet: 
J. Phys. Soc. Jpn. {\bf 80}, 013705 (2011).

\bibitem{Yel11}
E.~A. Yelland, J.~M. Barraclough, W.~Wang, K.~V. Kamenev and A.~D. Huxley:
Nature Phys. {\bf 7}, 890 (2011).

\bibitem{Mal12}
L.~Malone, L.~Howald, A.~Pourret, D.~Aoki, V.~Taufour, G.~Knebel and J.~Flouquet: 
Phys. Rev. B {\bf 85}, 024526 (2012).

\bibitem{Sam10}
M.~{Samsel-Czeka{\l}a}, S.~Elgazzar, P.~M. Oppeneer, E.~Talik, W.~Walerczyk and R.~Tro{\'c}: 
J. Phys.: Condens. Matter {\bf 22}, 015503 (2010).

\end{thebibliography}

\end{document}